\documentclass[journal]{IEEEtran}
\usepackage{pifont}
\usepackage{}
\usepackage[noadjust]{cite}
\usepackage{graphicx}
\usepackage{float}
\usepackage{amsmath}
\usepackage{enumerate}
\usepackage{amssymb}
\usepackage{fixltx2e}
\usepackage{color}
\usepackage{algorithm}
\usepackage{algorithmic}
\usepackage{amsmath}
\usepackage{multirow}
\usepackage{ragged2e}
\usepackage{booktabs}
\usepackage{subfigure}

\usepackage{xcolor,cite,etoolbox}
\makeatletter 
\pretocmd\@bibitem{\color{black}\csname keycolor#1\endcsname}{}{\fail}
\newcommand\citecolor[1]{\@namedef{keycolor#1}{\color{red}}}
\makeatother
\begin{document}
\pagenumbering{arabic}

\title{\LARGE{Mutual Information Rate of Gaussian and Truncated Gaussian Inputs on Intensity-Driven Signal Transduction Channels}}

\author{Xuan Chen, Fei Ji,~\emph{Member,~IEEE,} Miaowen Wen,~\emph{Senior Member,~IEEE,}\\
Yu Huang,~\emph{Member,~IEEE,}~Andrew W. Eckford,~\emph{Senior Member,~IEEE} \\

\thanks{

% Manuscript received April 21, 2022; revised June 22, 2022; accepted July 18, 2022. Date of publication XXXX XX, 2022;
% date of current version July 23, 2022. This work was supported in part by the National Natural Science Foundation of China under Grant 61871190 and in part by the Discovery Grant from the National Sciences and Engineering Council of Canada. The work of X. Chen is funded by China Scholarship Council No. 202006150127. \emph{(Corresponding author: Miaowen Wen.)}

Xuan Chen and Yu Huang are with the Research Center of Intelligent Communication Engineering, School of Electronics and Communication Engineering, Guangzhou University, Guangzhou 510006, China  (Email: eechenxuan@mail.scut.edu.cn, yuhuang@gzhu.edu.cn).

Fei Ji and Miaowen Wen are with the School of Electronic and Information Engineering, South China University of Technology, Guangzhou 510640, China (Email: \{eefeiji, eemwwen\}@scut.edu.cn). 
    
% Yu Huang is with the Research Center of Intelligent Communication Engineering, School of Electronics and Communication Engineering, Guangzhou University, Guangzhou 510006, China (Email: yuhuang@gzhu.edu.cn).
    
Andrew W. Eckford is with the Department of Electrical Engineering and Computer Science, York University, Toronto M3J 1P3, Canada (Email: aeckford@yorku.ca).

}}

%\thanks{This work has been funded in the framework of the IST project
%      IST-1999-12070 TRUST, which is partly funded by the European
%      Union. The authors would like to acknowledge the contributions
%      of their colleagues.}}
\maketitle
\vspace{-0.3cm}
\begin{abstract}

In this letter, we investigate the mutual information rate (MIR) achieved by an independent identically distributed (IID) Gaussian input on the intensity-driven signal transduction channel. Specifically, the asymptotic expression of the continuous-time MIR is given. Next, aiming at low computational complexity, we also deduce an approximately numerical solution for this MIR. Moreover, the corresponding lower and upper bounds that can be used to find the capacity-achieving input distribution parameters are derived in closed-form. Finally, simulation results show the accuracy of our analysis. 

\end{abstract}


\date{\today}
\renewcommand{\baselinestretch}{1.2}
% \thispagestyle{empty} \maketitle \thispagestyle{empty}
%\newpage
\setcounter{page}{1}

\IEEEpeerreviewmaketitle

\section{Introduction}

% introduce the signal transduction channel in molecular communication

Signal transduction, a typical form of molecular communication in nature, can support the communication between living cells. Examples of such systems include: binding of the acetylcholine (ACh) neurotransmitter to its receptor protein~\cite{ACh_receptor}, modulation of the channel opening transition by light intensity in the channelrhodopsin (ChR) protein \cite{ChR2_receptor}, and so~on.

At present, there are two main branches of signal transduction research in the communication field. One is to construct a stochastic model for the signal transduction system, while the other is to provide insights into these systems from the information-theoretic perspective. In the first branch, stochastic modeling of signal transduction as a communication channel has considered the chemical reactions in terms of Markov chains \cite{Model_signal_transduction_1} and in terms of the “noise” inherent in the binding process \cite{Model_signal_transduction_2}. Further, a linear noise approximation was developed for signal transduction channels in \cite{Model_signal_transduction_3}. In the second branch, Shannon capacity or mutual information of some typical signal transduction processes has been analyzed based on the above stochastic model. In particular, when considering a single receptor and multiple independent receptors, the Shannon capacity of two-state Markov signal transduction under arbitrary inputs was derived in \cite{Capacity_signal_transduction_3} and \cite{Capacity_signal_transduction_4}, respectively. Calculation of the mutual information rate (MIR) and capacity for individual receptors with ChR-2 (ChR2), ACh, and calmodulin (CaM) was performed~in~\cite{Capacity_signal_transduction_5}, where the states considered in the Markov chain are more diverse.

Notably, the previous work on the mutual information or channel capacity analysis of the signal transduction channel only considered the channel input to be independent and identically distributed (IID) discrete symbols, especially IID binary symbols \cite{Capacity_signal_transduction_3,Capacity_signal_transduction_4,Capacity_signal_transduction_5}.
However, when it comes to the input of the target system, continuous-valued variables are often used: the concentration of a ligand, the intensity of a light source, the potential difference across a cell membrane, etc.
In nature, the above input can be easily modeled as a Gaussian distribution, due to the central limit theorem. For example, the spatial intensity distribution for the LED (such as Laser TEM00 beam) can be approximated by a Gaussian profile ($>{95\%}$)~\cite{Gaussian_light}.
% Therefore, it is interesting to investigate the MIR of the signal transduction channel when considering the IID Gaussian continuous input. 
Against this background, in this paper, we propose to exploit the MIR achieved by an IID Gaussian or truncated Gaussian continuous input on the intensity-driven signal transduction channel. For ease of analysis, the approximate solution for the MIR is theoretically studied, and the corresponding bounds are also derived in closed-form. Finally, Monte Carlo simulations are carried out to verify the analysis. 

\section{System Model}

In this paper, we consider a typical signal transduction process, which can be regarded as a Markov chain model first, and then as a communication model. The complete equivalent process will be detailed in the sequel.

\subsection{Markov Chain Model of Signal Transduction}

The finite-state Markov chain model is first employed to describe the signal transduction process for a single receptor. Assuming a receptor with $k$ discrete states, there exists a $k$-dimensional vector of state occupancy probabilities ${\bf{p}}\left( t \right)$, i.e.,
\begin{align}\label{state occupancy probabilities}
{\bf{p}}\left( t \right) = \left[ {{p_1}\left( t \right),{p_2}\left( t \right), \ldots ,{p_k}\left( t \right)} \right],
\end{align}
where ${{p_i}\left( t \right)}$ denotes the probability that the receptor is in state $i$ at time $t$ with $i = 1,2, \ldots ,k$. It is well known that this probability evolves according to the master equation \cite{Capacity_signal_transduction_5}, i.e., 
\begin{align}\label{the master equation}
\frac{{d{\bf{p}}\left( t \right)}}{{dt}} = {\bf{p}}\left( t \right){\bf{Q}}\left( {x\left( t \right)} \right).
\end{align}
Here, ${\bf{Q}}\left( {x\left( t \right)} \right)$ is a $k \times k$ matrix of rate constants, where the element at the $i$-th row and $j$-th column of ${\bf{Q}}\left( {x\left( t \right)} \right)$, $q_{ij}\left( {x\left( t \right)} \right)$, is the instantaneous rate at which receptors starting~in state $i$ enter state $j$ under the input ${x\left( t \right)}$. Here, we define~a discrete time step as $\Delta t$ to analyze the Markov chain \cite{mathematic_expression}. When $\Delta t \to 0$, the master equation in \eqref{the master equation} can be approximated~as 
\newpage
\begin{align}
{\bf{p}}\left( {t + \Delta t} \right){\rm{ }} = {\bf{p}}\left( t \right)\left( {{\bf{I}} + {\bf{Q}}\Delta t} \right) + o\left( {\Delta t} \right),
\end{align}
where {\bf{I}} is the identity matrix and ${\bf{Q}}\left( {x\left( t \right)} \right)$ is simplified as ${\bf{Q}}$. Neglecting the high-order terms $ o\left( {\Delta t} \right)$, the channel state can be represented as a discrete-time Markov chain with the transition probability matrix
\begin{align}\label{transition probability matrix}
{\bf{P}} = {\bf{I}} +{\bf{Q}} \Delta t,
\end{align}
where ${\bf{P}}$ is given by using $p_{ij}$ as the $i$-th row and $j$-th column element, which indicates the probability of receptors moving from state $i$ to state $j$ in one time step. Note that ${\bf{P}}$ (and ${\bf{Q}}$) is dependent on $x\left( t \right)$, and thus the Markov chain described via ${\bf{P}}$ is not generally time-homogeneous if $x\left( t \right)$ is known.

\begin{figure}[t]
    \centering
        \includegraphics[width=1.5in]{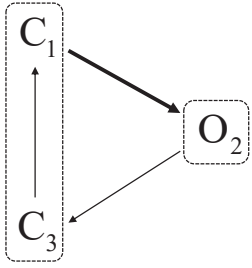}
    \caption{State transition diagram for ChR2. Sensitive transitions are depicted with bold arrows. States are labeled by their channel state: $\left\{ {{\rm{C}},{\rm{O}}} \right\}$ for closed and open, respectively; the state number is in subscript. Dashed lines surround all states in either the closed or open state.}
    \label{figure-State_transition}
\end{figure}

For clarity, we take the ChR2 receptor as an example to detail the state transition process. As shown in Fig.~\ref{figure-State_transition}, the ChR2 receptor has three states, namely ${\rm{C}}_1$, ${{\rm{O}}_2}$, and ${\rm{C}}_3$. Specifically, the ${{\rm{C}}_1} \to {{\rm{O}}_2}$ transition is sensitive to the input ${x\left( t \right)}$, while the ${{\rm{O}}_2} \to {{\rm{C}}_3}$ and ${{\rm{C}}_3} \to {{\rm{C}}_1}$ transitions~are insensitive. Here, the rate matrix for ChR2 can be written as
\begin{align}\label{rate matrix}
{\bf{Q}} = \bordermatrix{ %
 {\rm State} & {1} & {2} & {3}  \cr
 ~~{1}& R_{1}& q_{12}{x\left( t \right)} & 0\cr 
 ~~{2} & 0 &  R_{2} & q_{23}\cr
 ~~{3}& q_{31} & 0 & R_{3}\cr
},
\end{align}
where ${R_1} =  - {q_{12}}{x\left( t \right)}$, ${R_2} =  - {q_{23}}$, and ${R_3} =  - {q_{31}}$. By observing \eqref{rate matrix}, we define ${\mathcal S}$ (or ${\cal S}'$) as a set where the state transition is dependent (or independent) on ${x\left( t \right)}$. For the ChR2 receptor, we have ${\cal S} = \left\{ {{{\rm{C}}_1} \to {{\rm{C}}_1},{{\rm{C}}_1} \to {{\rm{O}}_2}} \right\}$, while ${\cal S}'$ 
includes all transitions except ${\mathcal S}$.

\subsection{Communication Model of Signal Transduction}

Here, the signal transduction will be regarded as a communication system with Markov channels, consisting of the input, output, and conditional input-output probability function.
\begin{itemize}
    \item \emph{Input}: The input $x\left( t \right)$ is the concentration or intensity in the environment at time $t$. When $x\left( t \right)$ is discretized in time, the input is $x\left( {i\Delta t} \right)$ for integers $i$. For clarity, let ${x_i} = x\left( {i\Delta t} \right)$ in the rest of this paper.   
    \item \emph{Output}: The output $y\left( t \right)$ means the state of receptors at time $t$. For simplicity, we employ the state number given by the subscript of the state label in Fig.~\ref{figure-State_transition} to mark the state of receptors. Therefore, for the ChR2 receptor, we have ${y\left( t \right)} =1,2,3$, corresponding to states $\left\{ {{{\text C}_1},{{\text O}_2},{{\text C}_3}} \right\}$, respectively. Similarly, $y(t)$ can be discretized in time as ${y_i} = y\left( {i\Delta t} \right)$. 
    % \item \emph{Input-output Relationship}. As a Markov channel, ${y_i}$, the~state of the receptor at time $i$, depends on the current input ${x_i}$ and the previous channel output ${y_{i-1}}$. Here, we can write the conditional input-output probability as $p({\bf{y}}\mid {\bf{x}}) = \prod\limits_{i = 1}^n p \left( {{y_i}\mid {x_i},{y_{i - 1}}} \right)$, where $p\left( {{y_1}\mid {x_1},{y_0}} \right) = p\left( {{y_1}\mid {x_1}} \right)$, ${\bf{x}} = \left[ {{x_1},{x_2}, \ldots {x_n}} \right]$, and ${\bf{y}} = \left[ {{y_1},{y_2}, \ldots {y_n}} \right]$.  From~\eqref{transition probability matrix}, we have $p\left( {{y_i}\mid {x_i},{y_{i - 1}}} \right) = {p_{{y_{i - 1}},{y_i}}}$.
    \item \emph{Input-output relationship}: As a Markov channel, ${y_i}$, the~state of the receptor at time $i$, depends on the current input ${x_i}$ and the previous channel output ${y_{i-1}}$. Here, we can write the conditional input-output probability as 
    \begin{align}\label{conditional_PMF}
    {p_{Y_1^n\left| {X_1^n} \right.}}\left( {y_1^n\left| {x_1^n} \right.} \right) = \prod\limits_{i = 1}^n {{p_{{Y_i}\left| {{X_i},{Y_{i - 1}}} \right.}}\left( {{y_i}\left| {{x_i},{y_{i - 1}}} \right.} \right)} ,
    \end{align}
    where ${x_1^n} = \left[ {{x_1},{x_2}, \ldots, {x_n}} \right]$, ${y_1^n}= \left[ {{y_1},{y_2}, \ldots, {y_n}} \right]$, and ${p_{{Y_1}\left| {{X_1},{Y_0}} \right.}}\left( {{y_1}\left| {{x_1},{y_0}} \right.} \right) = {p_{{Y_1}\left| {{X_1}} \right.}}\left( {{y_1}\left| {{x_1}} \right.} \right)$. From~\eqref{transition probability matrix}, we can further obtain ${p_{{Y_i}\left| {{X_i},{Y_{i - 1}}} \right.}}\left( {{y_i}\left| {{x_i},{y_{i - 1}}} \right.} \right) = {p_{{y_{i - 1}}{y_i}}}$.
\end{itemize}
In particular, we will omit the subscripts for probability functions where unambiguous, e.g., ${p_Y}\left( y \right)$ becomes ${p}\left( y \right)$. 

Additionally, we assume that the input $x_i$ follows an IID Gaussian distribution, and then $x_i$ can be simplified as $x$. In this work, the input $x$ denotes the intensity, which should be in a certain range without reaching infinity. To the best of our knowledge, the truncated Gaussian distribution, in which the range of definition is made finite at both ends of the interval, is employed to avoid extreme values. Therefore, we suppose that $x$ has a Gaussian distribution with mean ${\bar \mu}$ and variance ${\bar \sigma} ^{2}$ and lies within the interval $[ {a,b}]$ with $ 0 \le a < b <  + \infty $. Then $x$ conditional on $[ {a,b}]$ obeys the truncated~Gaussian~distribution, whose probability density function can be given by \cite{Truncated_Gaussian_1}
\begin{align}
p(x)= \left\{ \hspace{-0.2cm}{\begin{array}{*{20}{c}}
0 \hfill & {{\rm{ if }}~x < a} \hfill  \\
{\frac{{\phi \left( {{\bar \mu} ,{{\bar \sigma} ^2};x} \right)}}{{\Phi \left( {{\bar \mu} ,{{\bar \sigma} ^2};b} \right) - \Phi \left( {{\bar \mu} ,{{\bar \sigma} ^2};x} \right)}}} \hfill & {{\rm{ if }}~a \le x \le b} \hfill  \\
0 \hfill & {{\rm{ if }}~x > b} \hfill  \\
\end{array}} \right.,
\end{align}
where 
\begin{align}
\begin{array}{c}
\phi \left( {\bar \mu ,{{\bar \sigma }^2};x} \right) = \frac{1}{{\sqrt {2\pi {{\bar \sigma }^2}} }}{e^{\left( { - \frac{{{{\left( {x - \bar \mu } \right)}^2}}}{{2{{\bar \sigma }^2}}}} \right)}} \\ 
\\ \vspace{-0.55cm} \\
\Phi \left( {\bar \mu ,{{\bar \sigma }^2};b} \right) = \frac{1}{2}\left( {1 + {\rm{ erf}}\left( {\left( {x - \bar \mu } \right)/\sqrt {2{{\bar \sigma }^2}} } \right)} \right) \\
 \end{array}, \nonumber
\end{align}
where ${\rm{erf}}( \cdot )$ is the error function. For the considered $x$, its mean and variance will be updated~as
\begin{align}
\begin{array}{c}
 \mu  = \bar \mu  - \frac{{\phi (0,1;\beta ) - \phi (0,1;\alpha )}}{{\Phi (0,1;\beta ) - \Phi (0,1;\alpha )}}\bar \sigma  \\
\\ \vspace{-0.66cm} \\
 {\sigma ^2} = {{\bar \sigma }^2}\left\{ {1 - \frac{{\beta \phi (0,1;\beta ) - \alpha \phi (0,1;\alpha )}}{{\Phi (0,1;\beta ) - \Phi (0,1;\alpha )}} - {{\left( {\frac{{\phi (0,1;\beta ) - \phi (0,1;\alpha )}}{{\Phi (0,1;\beta ) - \Phi (0,1;\alpha )}}} \right)}^2}} \right\} \\ 
 \end{array},  \nonumber
\end{align}
where $\alpha  = \frac{{a - {\bar \mu} }}{{\bar \sigma} }$ and $\beta  = \frac{{b - {\bar \mu} }}{{\bar \sigma} }$. Under the above assumption, the receptor states ${Y_1^n}$ can form a time-homogeneous Markov chain \cite{mathematic_expression}, i.e.,
\begin{align}\label{IID_proba_y}
 p\left( {y_1^n} \right) = \prod\limits_{i = 1}^n {\int\limits_x {p\left( x \right)p\left( {{y_i}\left| {x,{y_{i - 1}}} \right.} \right)dx} } = \prod\limits_{i = 1}^n {{{\bar p}_{{y_{i - 1}}{y_i}}}},
\end{align}
where ${{{\bar p}_{{y_{i - 1}}{y_i}}}}$ is the ${y_{i - 1}}$-th row and ${y_{i}}$-th column element of ${\bf{\bar P}}$ and ${\bf{\bar P}}$ is the transition probability matrix of ${Y_1^n}$, written as
\begin{align}\label{average transition probability matrix}
{\bf{\bar P}} = E\left[ {\bf{P}} \right] = {\bf{I}} + E\left[ {\bf{Q}} \right]\Delta t.
\end{align}
Recalling \eqref{transition probability matrix} and \eqref{rate matrix}, we can replace $x$ (or $x\left(t\right)$) in these terms with  $E\left[ x\right]$ to form $ E\left[ {\bf{ P}} \right]$ and $ E\left[ {\bf{Q}} \right]$, respectively, since the sensitive terms in ${\bf{ P}}$ and ${\bf{Q}}$ are assumed to be linear in $x$.

\section{MIR of Signal Transduction}\label{derivation_for_MIR}

In this section, we first give an asymptotic expression for the MIR in continuous time, i.e., obtaining $\mathop {\lim }\limits_{\Delta t \to 0} \frac{{{\cal I}\left( {X;Y} \right)}}{{\Delta t}}$. Next, we derive an approximately numerical solution for the continuous-time MIR. Finally, its lower and upper bounds are deduced.

\subsection{Calculation for the Continuous-time MIR}

For any communication system with input $x$ and output~$y_i$, the MIR can be given by \cite{Capacity_signal_transduction_5}
\begin{align}\label{capacity_calculation}
\hspace{-0.25cm}\text{MIR} &= {\textstyle{{{\cal I}\left( {X;Y} \right)} \over {\Delta t}}} \nonumber\\
&= \mathop {\lim }\limits_{n \to \infty } \frac{1}{n{\Delta t}}I\left( {Y_1^n;X_1^n} \right) \nonumber\\
&= \mathop {\lim }\limits_{n \to \infty } \frac{1}{\Delta t}\left\{ {H\left( {{Y_n}\left| {{Y_{n - 1}}} \right.} \right) - H\left( {{Y_n}\left| {{X_n},{Y_{n - 1}}} \right.} \right)} \right\}.
\end{align}
Note that ${\cal I}\left( {X;Y} \right) $ in \eqref{capacity_calculation} is the mutual information exchanged between $X$ and $Y$ per channel use with a duration $\Delta t$. Correspondingly, the unit of the MIR is bits/s, when ${\log _2}\left(  \cdot  \right)$ is used for the entropy calculation. 
With the aid of \eqref{IID_proba_y}, we~have
\begin{align}\label{entropy_for_output}
\hspace{-0.2cm}H\left( {{Y_n}\left| {{Y_{n - 1}}} \right.} \right) &=  - E\left[ {\log p\left( {{y_n}|{y_{n - 1}}} \right)} \right] \nonumber\\ 
&=  - \sum\limits_{\left\{ {{y_{n - 1}},{y_n}} \right\}} {{\pi _{{y_{n - 1}}}}} {{\bar p}_{{y_{n - 1}}{y_n}}}\log{{\bar p}_{{y_{n - 1}}{y_n}}},
\end{align}
and
\begin{align}\label{entropy_for_output_condition_x}
&H\left( {{Y_n}\left| {{X_n},{Y_{n - 1}}} \right.} \right) \nonumber\\
&= \int_x {p\left( x \right)H\left( {{Y_n}\left| {{X_n} = x,{Y_{n - 1}}} \right.} \right)dx}  \nonumber\\ 
&=  - \int_x {p\left( x \right)E\left[ {\log p\left( {{y_n}|x,{y_{n - 1}}} \right)} \right]dx}  \nonumber\\ 
&=  - \sum\limits_{\left\{ {{y_{n - 1}},{y_n}} \right\} \in {\cal S}} {{\pi _{{y_{n - 1}}}}\int_x {p\left( x \right)} {p_{{y_{n - 1}}{y_n}}}\log {p_{{y_{n - 1}}{y_n}}}} dx  \nonumber\\
&~~~-\sum\limits_{\left\{ {{y_{n - 1}},{y_n}} \right\} \in {\cal S}'} {{\pi _{{y_{n - 1}}}}{p_{{y_{n - 1}}{y_n}}}\log {p_{{y_{n - 1}}{y_n}}}} ,
\end{align}
where ${{\pi _{{y_{n - 1}}}}}$ is the steady-state marginal probability that the receptor is in state ${{y_{n - 1}}}$, which is the solution to the following system of equations:
\begin{align}\label{calculation for pi}
\left\{ \begin{array}{l}
 {\bf{\pi \bar P}} = {\bf{\pi}} \\ 
 \\\vspace{-0.77cm} \\
 \sum\limits_{{y_{n - 1}}} {{\pi _{{y_{n - 1}}}}}  = 1 \\ 
 \end{array} \right. . 
\end{align}
It is clear from \eqref{average transition probability matrix} and \eqref{calculation for pi} that ${{\pi _{{y_{n - 1}}}}}$ is only dependent on $E\left[x\right]$ and $q_{ij}$. Besides, since the transition in ${\cal S}'$ is independent on $x$, we have ${\bar p}_{{y_{n - 1}}{y_n}} = {p}_{{y_{n - 1}}{y_n}} $ for ${\left\{ {{y_{n-1}},{y_{n}}} \right\} \in {\cal S}'}$. Here, substituting \eqref{entropy_for_output} and \eqref{entropy_for_output_condition_x} into \eqref{capacity_calculation} yields 
\begin{align}\label{MIR_discrete} 
{\textstyle{{{\cal I}\left( {X;Y} \right)} \over {\Delta t}}}  =& \frac{1}{\Delta t}{\sum _{\left( {{y_{n - 1}},{y_n}} \right) \in {\cal S}}}{\pi _{{y_{n - 1}}}}\left( \int_{{x}} {p\left( {{x}} \right)\phi \left( {{p}_{{y_{n- 1}}{y_n}}} \right)d{x}} \right. \nonumber\\ 
 &\left.- \phi \left( {\int_{{x}} {p\left( {{x}} \right){p}_{{y_{n - 1}}{y_n}}d{x}} } \right) \right),
\end{align}
and
\begin{align}
\phi (p) = \left\{ {\begin{array}{*{20}{c}}
   {0,} \hfill & {p = 0} \hfill  \\
   {p\log p,} \hfill & {p \ne 0} \hfill  \\
\end{array}} \right. .
\end{align}
Further, we will compute an asymptotic expression for the MIR in continuous time, i.e., $\mathop {\lim }\limits_{\Delta t \to 0} {\textstyle{{{\cal I}\left( {X;Y} \right)} \over {\Delta t}}}$.
% Next, we will compute the continuous time limit of the discrete-time MIR, i.e., the limit of ${{\cal I}\left( {X;Y} \right)/{\Delta t}}$ as $\Delta t \to 0$. 
First, we have 
\newpage
\begin{align}\label{limit_MI}
\mathop {\lim }\limits_{\Delta t \to 0} \frac{{{\cal I}(X;Y)}}{{\Delta t}} =& \sum\limits_{\left\{ {{y_{n-1}},{y_{n}}} \right\} \in {\cal S},{y_{n-1}} \ne {y_{n}}} {\iota_{\left( {{y_{n - 1}},{y_n}} \right)}}  \nonumber\\
&+ \sum\limits_{\left\{ {{y_{n-1}},{y_{n}}} \right\} \in {\cal S},{y_{n-1}} = {y_{n}}} {\iota_{\left( {{y_{n-1}},{y_n}} \right)}} ,
\end{align}
where 
\begin{align}\label{limit_single}
{\iota _{\left( {{y_{n - 1}},{y_n}} \right)}} =& \mathop {\lim }\limits_{\Delta t \to 0} {\pi _{{y_{n - 1}}}}\left( {\frac{{\int_x {p\left( x \right)\phi \left( {{p_{{y_{n - 1}}{y_n}}}} \right)dx} }}{{\Delta t}}} \right. \nonumber\\
&\left. { - \frac{{\phi \left( {\int_x {p\left( x \right){p_{{y_{n - 1}}{y_n}}}dx} } \right)}}{{\Delta t}}} \right).
\end{align}
By using the L'Hôpital's rule for \eqref{limit_single}, we find $\iota_{\left( {{y_{n- 1}},{y_n}} \right)} = 0$ when $\left\{ {{y_{n-1}},{y_{n}}} \right\} \in {\cal S}$ and ${y_{n-1}} = {y_{n}}$ \cite{mathematic_expression}. Hence, \eqref{limit_MI} can be further re-written as
\begin{align}\label{limit_MI_simplified_without_i}
\hspace{-0.3cm}\mathop {\lim }\limits_{\Delta t \to 0}\hspace{-0.15cm}\frac{{{\cal I}(X;Y)}}{{\Delta t}} \hspace{-0.05cm}=\hspace{-0.05cm} {g_{{y_{n-1}},{y_{n}}}}\hspace{-0.1cm}\left(\hspace{-0.05cm} \int_x\hspace{-0.1cm}{p\left( x \right)x\ln \left( x \right)dx} \hspace{-0.05cm}- \hspace{-0.05cm}{ \mu}\ln {{\mu}} \hspace{-0.05cm}\right),
\end{align}
where 
\begin{align}
{g_{{y_{n-1}},{y_{n}}}} = \sum\limits_{\left\{ {{y_{n-1}},{y_{n}}} \right\} \in {\cal S},{y_{n-1}} \ne {y_{n}}} {\frac{{{\pi _{{y_{n - 1}}}}{q_{{y_{n - 1}}{y_n}}}}}{{\ln 2}}} .
\end{align}
So far, the derivation for $\mathop {\lim }\limits_{\Delta t \to 0}\hspace{-0.15cm}\frac{{{\cal I}(X;Y)}}{{\Delta t}}$ has been finished. 

\subsection{Approximately Numerical Solution for $\mathop {\lim }\limits_{\Delta t \to 0} \frac{{{\cal I}(X;Y)}}{{\Delta t}}$}\label{derivation_for_numerical_solution}

In this subsection, we provide an approximately numerical~solution for \eqref{limit_MI_simplified_without_i}. By observing \eqref{limit_MI_simplified_without_i}, we need to focus~on the item with $\int_x {p\left( x \right)x\ln\left( { x} \right)dx}$, since the other items in \eqref{limit_MI_simplified_without_i} are given for the target system. According to the McLaughlin formula, $\ln \left( {{x}} \right)$ can be approximately expressed~as \cite{mathematic_expression}
\begin{align}\label{lnx_approximation}
 \ln \left( {{x}} \right) = \ln \left( {1 + \left( {x - 1} \right)} \right) = \sum\limits_{k = 1}^\infty  {{{\left( { - 1} \right)}^{k - 1}}\frac{{{{\left( {x - 1} \right)}^k}}}{k}} .  
\end{align}
It is worth noting that when $0 < x \le 2$, the McLaughlin series of $\ln \left( {1 + \left( {x - 1} \right)} \right)$ holds and successfully converges to $\ln \left( {1 + \left( {x - 1} \right)} \right)$. Therefore, we assume $0 < x \le 2$ in this paper.\footnote{This assumption can easily hold for the considered system. For a truncated Gaussian random variable, its product with a constant still obeys a truncated Gaussian distribution. For example, when $x \sim {\cal N}(\mu ,{\sigma ^2}|x \in [a,b])$, we have $qx \sim {\cal N}(q\mu ,{(q\sigma )^2}|qx \in [qa,qb])$ for any possible $q$ \cite{Truncated_Gaussian_1}. Herein, the value range of $x$ can be adjusted via $q$.}Substituting \eqref{lnx_approximation} into $\int_x {p\left( x \right)x\ln\left( { x} \right)dx}$, we can have 
\begin{align}\label{xlnx_1}
&\int\limits_x {p\left( x \right)x\left( {\ln \left( {{x}} \right)} \right)dx} \nonumber\\ 
&= \sum\limits_{k = 1}^\infty  {\frac{{{{\left( { - 1} \right)}^{k - 1}}}}{k}} \int\limits_x {p\left( x \right)x{{\left( {x - 1} \right)}^k}dx} \nonumber\\ 
&=  - 1 + E\left[ x \right] + \sum\limits_{k = 2}^\infty  {\frac{{{{\left( { - 1} \right)}^k}}}{{k\left( {k - 1} \right)}}} E\left[ {{{\left( {x-1} \right)}^k}} \right] \nonumber\\
&=  - 1 + E\left[ x \right] + \sum\limits_{k = 2}^\infty  {\sum\limits_{m = 0}^k {\frac{{{{\left( { - 1} \right)}^{m}}C\left( {k,m} \right)}}{{k\left( {k - 1} \right)}}E\left[ {{x^m}} \right]} } ,
\end{align}
where $C\left( { \cdot , \cdot } \right)$ is the binomial coefficient and ${E\left[ {{x^m}} \right]}$ denotes the $m$-th moment. According to the description in \cite{Truncated_Gaussian_1}, ${E\left[ {{x^m}} \right]}$ for the truncated Gaussian distribution can be calculated as:
\newpage
% \begin{align}
% \begin{array}{l}
%  E\left[ {{x^n}} \right] = \sum\limits_{i = 0}^n {\left( {\begin{array}{*{20}{c}}
%   n  \\
%   i  \\
% \end{array}} \right)} {{\bar \sigma} ^i}{{\bar \mu} ^{k - i}}{L_i} \\ 
%  {L_0} = 1 \\ 
%  {L_1} =  - \frac{{\phi (0,1;\beta ) - \phi (0,1;\alpha )}}{{\Phi (0,1;\beta ) - \Phi (0,1;\alpha )}} \\ 
%  {L_i} =  - \frac{{{\beta ^{i - 1}}\phi (0,1;\beta ) - {\alpha ^{i - 1}}\phi (0,1;\alpha )}}{{\Phi (0,1;\beta ) - \Phi (0,1;\alpha )}} + (i - 1){L_{i - 2}} \\ 
%  \end{array}
% \end{align}
\begin{align}\label{Moment_truncated_Gaussian}
E\left[ {{x^m}} \right] = \sum\limits_{i = 0}^m C\left( {m,i} \right) {{\bar \sigma} ^i}{{\bar \mu} ^{m - i}}{L_i},~~~~~~~~~~~~~~~~~\\
\begin{array}{l}
 {L_0} = 1 \\ 
 {L_1} =  - \frac{{\phi (0,1;\beta ) - \phi (0,1;\alpha )}}{{\Phi (0,1;\beta ) - \Phi (0,1;\alpha )}} \\ 
 {L_i} =  - \frac{{{\beta ^{i - 1}}\phi (0,1;\beta ) - {\alpha ^{i - 1}}\phi (0,1;\alpha )}}{{\Phi (0,1;\beta ) - \Phi (0,1;\alpha )}} + (i - 1){L_{i - 2}} \\
 \end{array} .
\end{align}
Finally, the limit of the continuous-time MIR in \eqref{limit_MI_simplified_without_i} can be numerically expressed as
\begin{align}\label{Approximately Numerical Solution}
\mathop {\lim }\limits_{\Delta t \to 0} \frac{{{\cal I}(X;Y)}}{{\Delta t}}= &{g_{{y_{n-1}},{y_{n}}}} \left(   \sum\limits_{k = 2}^\infty  {\sum\limits_{m = 0}^k {\frac{{{{\left( { - 1} \right)}^m}C\left( {k,m} \right)}}{{k\left( {k - 1} \right)}}E\left[ {{x^m}} \right]} } \right.\nonumber\\
&\left.- E\left[ x \right]\ln \frac{{E\left[ x \right]}}{e} - 1 \right) .
\end{align}

\subsection{Bounds for $\mathop {\lim }\limits_{\Delta t \to 0} \frac{{{\cal I}(X;Y)}}{{\Delta t}}$}\label{derivation_for_bounds}

In this subsection, we will provide lower and upper bounds to estimate the possible range of the continuous-time MIR.  
It is clear from \eqref{limit_MI_simplified_without_i} that $\left( \int_{{x}} {p\left( {{x}} \right){x}\ln \left( {{x}} \right)d{x}}\hspace{-0.1cm}-\hspace{-0.1cm}{ \mu} \ln {{ \mu} } \right)$ is a Jensen gap of ${x}\ln \left( {{x}} \right)$. According to the description in \cite{Bounds_of_capacity_tight}, we can calculate a bound for the considered Jensen gap. First, we define $f\left( x \right) = x\ln \left( {{x}} \right)$. Next, assuming that $E{\left| {x - \mu } \right|^s} < \infty$ holds for $s = 2m,m = 1,2,3 \ldots$ and $f\left( x \right)$ is a $\left(s+1\right)$-times differentiable function on $\forall x \in \left( {a,b} \right)$, we further have 
\begin{align}
{h^{(s)}}(x;\mu ) = \frac{{f(x) - f(\mu )}}{{{{(x - \mu )}^s}}} - \sum\limits_{i = 1}^{s - 1} {\frac{{{f^{(i)}}(\mu )}}{{i!{{(x - \mu )}^{s - i}}}}} ,
\end{align}
where ${f^{(i)}}(x) = \frac{{{{\rm{d}}^i}}}{{{\rm{d}}{x^i}}}f(x)$.
Based on Theorem 2.1 in \cite{Bounds_of_capacity_tight}, we have
\begin{align}\label{Jensen_gap_1}
\hspace{-0.2cm}\begin{array}{l}
 E[f\left( x \right)] \hspace{-0.08cm}-\hspace{-0.08cm} f\left( {E\left[ x \right]} \right) \hspace{-0.08cm}\ge\hspace{-0.08cm} \sum\limits_{i = 1}^{s - 1} {r_i}
 + \mathop {\inf }\limits_{x \in (a,b)} \{ {h^{(s)}}(x;\mu )\} {\mu _s} \\ 
 E[f\left( x \right)] \hspace{-0.08cm}-\hspace{-0.08cm} f\left( {E\left[ x \right]} \right) \hspace{-0.08cm}\le\hspace{-0.08cm} \sum\limits_{i = 1}^{s - 1} {r_i} + \mathop {\sup }\limits_{x \in (a,b)} \{ {h^{(s)}}(x;\mu )\} {\mu _s} \\ 
 \end{array},
\end{align}
where ${r_i} = \frac{{{\mu _i}}}{{i!}}{f^{(i)}}(\mu )$ and ${\mu _i} = E{\left[ {x - \mu } \right]^i}$ is the $i$-th central moment of $x$ with $i = 1,2, \ldots ,s$, which can be solved via \eqref{Moment_truncated_Gaussian}. Besides, it is obvious from Lemma 2.3 of~\cite{Bounds_of_capacity_tight} that if ${f^{(s-1)}}(x)$ is strictly convex (concave), ${h^{(s)}}(x;\mu )$ strictly increases (decreases) with respect to $x$. Assuming that ${f^{(s-1)}}(x)$ is concave for $\forall x \in \left( {a,b} \right)$, \eqref{Jensen_gap_1} can be re-written~as
\begin{align} 
\begin{array}{l}
 E[f\left( x \right)] - f\left( {E\left[ x \right]} \right) \ge \sum\limits_{i = 1}^{s - 1} {{r^i}\left( \mu  \right)}  + {h^{(s)}}(b;\mu){\mu _s} \\ 
 E[f\left( x \right)] - f\left( {E\left[ x \right]} \right) \le \sum\limits_{i = 1}^{s - 1} {{r^i}\left( \mu  \right)}  + {h^{(s)}}(a;\mu){\mu _s} \\ 
 \end{array}.
\end{align}
Following the example given in \cite{Bounds_of_capacity_tight}, in this paper, we consider $s=2$ and $s=4$ to derive the bounds for the target MIR.

\subsubsection{$s=2$} From the second derivative of $f^{(1)}\left( x \right)$ with respect to $x$, i.e., ${f^{(3)}}(x)=  - \frac{1}{{{x^2}}}$, it is clear that $f^{(1)}\left( x \right)$ is concave. Therefore, $h^{(2)}\left( {x;{ \mu} } \right) $ is monotonically decreasing with respect to $x$. Based on the aforementioned, \eqref{Jensen_gap_1} can be rewritten as
\begin{align}\label{Jensen_gap_s_2}
\begin{array}{l}
\hspace{-0.15cm}E[f\left( x \right)] \hspace{-0.05cm}-\hspace{-0.05cm} f\left( {E\left[ x \right]} \right) \hspace{-0.05cm}\ge\hspace{-0.05cm} \left\{ {\frac{{b\ln b - u\ln u}}{{{{\left( {b - u} \right)}^2}}} \hspace{-0.05cm}-\hspace{-0.05cm} \frac{{1 + \ln u}}{{b - u}}} \right\}{{ \sigma}}^2 \\ 
\hspace{-0.15cm}E[f\left( x \right)] \hspace{-0.05cm}-\hspace{-0.05cm} f\left( {E\left[ x \right]} \right) \hspace{-0.05cm}\le\hspace{-0.05cm} \left\{ {\frac{{a\ln a - u\ln u}}{{{{\left( {a - u} \right)}^2}}} \hspace{-0.05cm}-\hspace{-0.05cm} \frac{{1 + \ln u}}{{a - u}}} \right\}{{ \sigma}}^2 \\ 
 \end{array}.
\end{align}
Substituting \eqref{Jensen_gap_s_2} into 
\eqref{limit_MI_simplified_without_i} yields
\begin{align}\label{bound_July}
\begin{array}{l}
\hspace{-0.35cm}\mathop {\lim }\limits_{\Delta t \to 0} \frac{{{\cal I}(X;Y)}}{{\Delta t}} \hspace{-0.05cm}\ge\hspace{-0.05cm} {g_{{{y_{n-1}},{y_{n}}}}}\left\{ {\frac{{b\ln b - u\ln u}}{{{{\left( {b - u} \right)}^2}}} \hspace{-0.05cm}-\hspace{-0.05cm} \frac{{1 + \ln u}}{{b - u}}} \right\}\hspace{-0.05cm}{{ \sigma}}^2 \\ 
\hspace{-0.35cm}\mathop {\lim }\limits_{\Delta t \to 0} \frac{{{\cal I}(X;Y)}}{{\Delta t}} \hspace{-0.05cm}\le\hspace{-0.05cm} {g_{{{y_{n-1}},{y_{n}}}}}\left\{ {\frac{{a\ln a - u\ln u}}{{{{\left( {a - u} \right)}^2}}} \hspace{-0.05cm}-\hspace{-0.05cm} \frac{{1 + \ln u}}{{a - u}}} \right\}\hspace{-0.05cm}{{ \sigma}}^2 \\ 
 \end{array}.
\end{align}

\subsubsection{$s=4$}
Similarly, $h^{(4)}\left( {x;{ \mu} } \right) $ is monotonically decreasing with respect to $x$, due to ${f^{(3)}}(x) =  - \frac{6}{{{x^4}}} < 0$ for $\forall x \in \left( {a,b} \right)$. Here, \eqref{Jensen_gap_1} can be rewritten as 
\begin{align}\label{Jensen_gap_s_4}
\begin{array}{l}
 E[f\left( x \right)] - f\left( {E\left[ x \right]} \right) \ge \frac{{{\sigma ^2}}}{{2\mu }} - \frac{{{\mu _3}}}{{6{\mu ^2}}} + {h^{(4)}}(b;\mu ){\mu _4} \\ 
 E[f\left( x \right)] - f\left( {E\left[ x \right]} \right) \le \frac{{{\sigma ^2}}}{{2\mu }} - \frac{{{\mu _3}}}{{6{\mu ^2}}} + {h^{(4)}}(a;\mu ){\mu _4} \\ 
 \end{array}.
\end{align}
Here, the bounds for $\mathop {\lim }\limits_{\Delta t \to 0} \frac{{{\cal I}(X;Y)}}{{\Delta t}}$ can be obtained by substituting \eqref{Jensen_gap_s_4} into \eqref{limit_MI_simplified_without_i}.

\emph{Remark:} Note that the derivation in Section~\ref{derivation_for_MIR} is valid for all ${\bf{Q}}$, when the IID input $x$ follows any possible distribution within the range of~$\left( {0,2} \right]$. Therefore, we conclude that all MIR expressions derived in this paper can be extended to arbitrary signal transduction systems under the above condition.
 
\section{Numerical Results and Analysis}

In this section, we employ the ChR2 receptor and the ACh receptor as examples in our analysis, where their parameters have been listed in Table~I and Table~II of \cite{Capacity_signal_transduction_5}, respectively. Moreover, for analysis, we adjust the input range for these two receptors as $x \in \left[ {10^{-5},2} \right]$ and $x \in \left[ {2 \times 10^{-2},2} \right]$.

\begin{figure*}[htb]
    \centering
       \subfigure[${\bar \sigma}= 0.5$]{
	\includegraphics[width=3in]{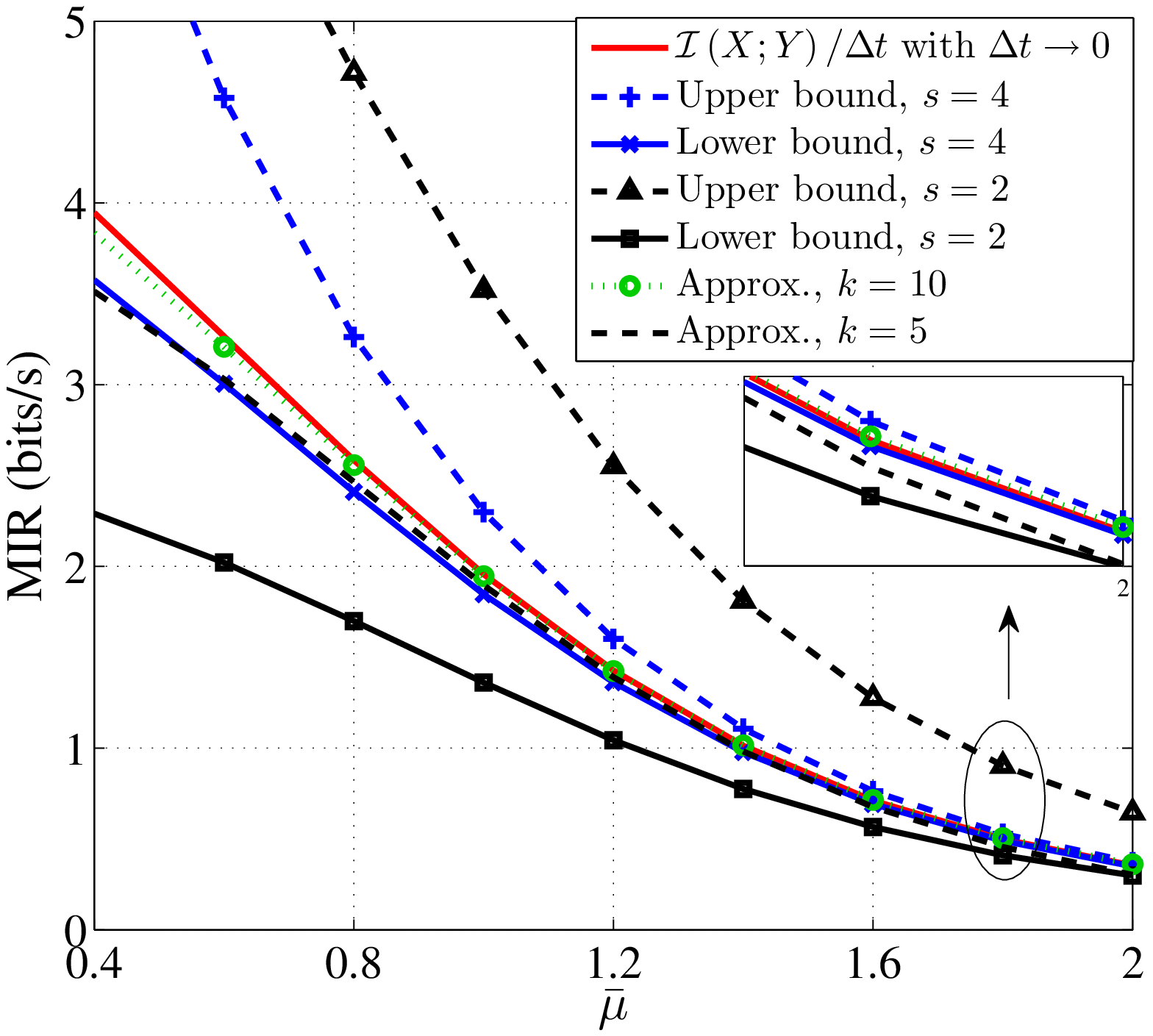}
	\label{expectation_change_ChR2}
	}
	    \subfigure[${\bar \mu} = 1$]{
        \includegraphics[width=3in]{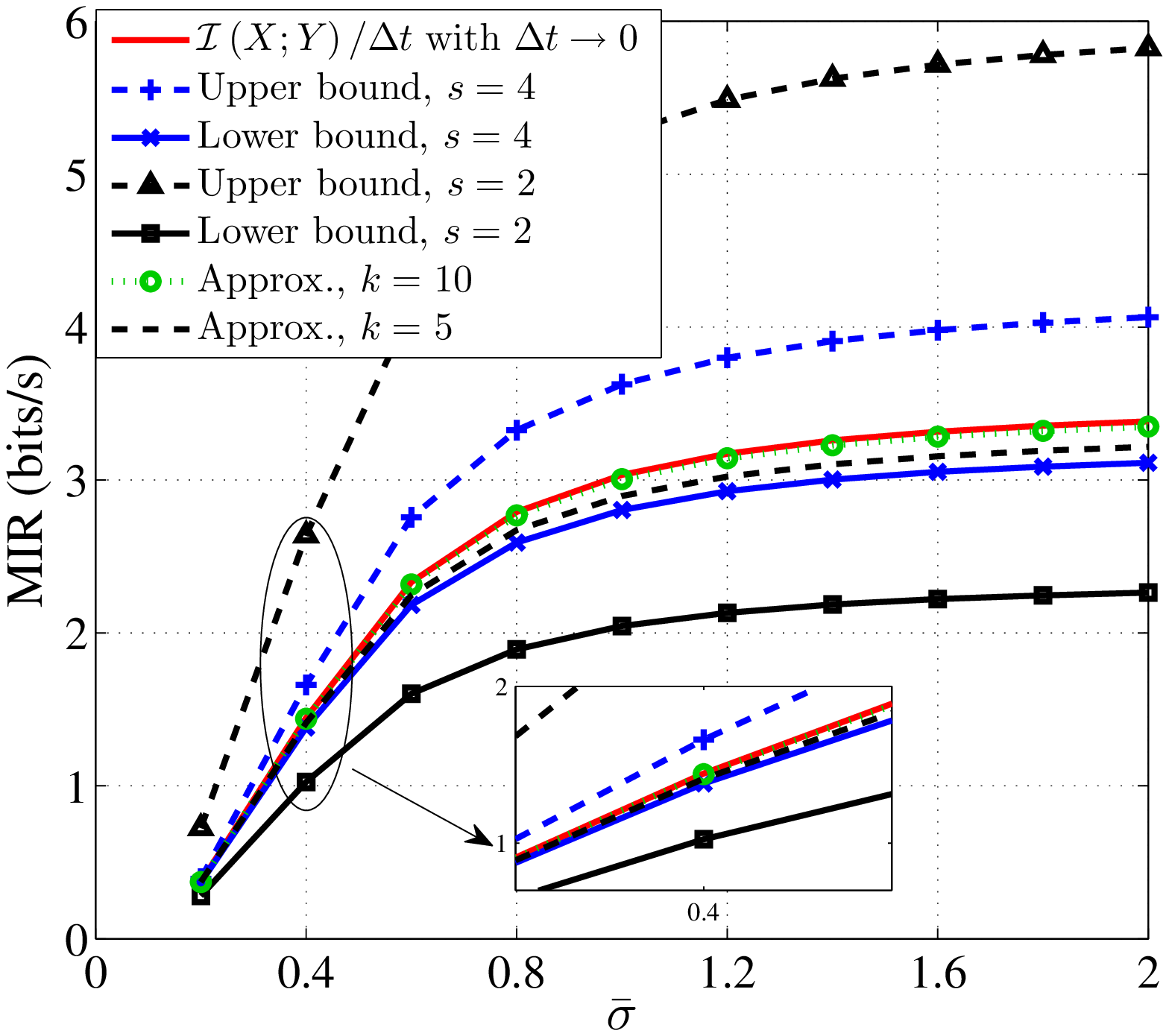}
    \label{s_var_change_ChR2}
    }
    \vspace{-0.25cm}
    \caption{Numerical verification for the ChR2 receptor: the approximately numerical solution and bounds for $\mathop {\lim }\limits_{\Delta t \to 0} {\textstyle{{{\cal I}\left( {X;Y} \right)} \over {\Delta t}}}$, where it is assumed that the input $x$ follows the truncated Gaussian distribution with $x \in \left[ {10^{-5},2} \right]$.}
    \label{Numerical verification for the ChR2 receptor}
\end{figure*}

Fig.~\ref{Numerical verification for the ChR2 receptor} shows the continuous-time MIR as well as the corresponding numerical solution and bounds for the ChR2 receptor. For clarity, we employ $\bar \mu$ and $\bar \sigma$ as variables to study the MIR when defining $x \sim {\cal N}\left( {\bar \mu ,{{\bar \sigma }^2}} \right)$ and $x\left| {x \in \left[ {{{10}^{ - 5}},2} \right]} \right. \sim {\cal N}\left( {\mu ,{\sigma ^2}} \right)$. One can easily observe from Fig.~\ref{Numerical verification for the ChR2 receptor} that the~MIR can achieve good performance when $x$ with a small value holds a sizable proportion of all inputs, corroborating the results described in \cite{Capacity_signal_transduction_5}. Further, it can be determined from Fig.~\ref{Numerical verification for the ChR2 receptor} that the MIR and its approximately numerical counterparts can accurately match when $x$ is concentrated around large values, and the accuracy performance is proportional~to~$k$. 
Moreover, we can find that the bounds are gradually tighter as $s$ goes large, while the calculation complexity is also increasing. Specifically, when $s =4$, the derived bounds can provide a relatively narrow range to estimate the exact MIR.
Similar numerical results are attained for the ACh receptor, when varying $\bar \mu$ or $\bar \sigma$.
In summary, the derived numerical solution and bounds for the MIR can give an accurate approximation with low complexity, when obtaining the exact MIR is challenging.

\begin{figure*}[htb]
    \centering
    \subfigure[MIR]{
        \includegraphics[width=2.2in]{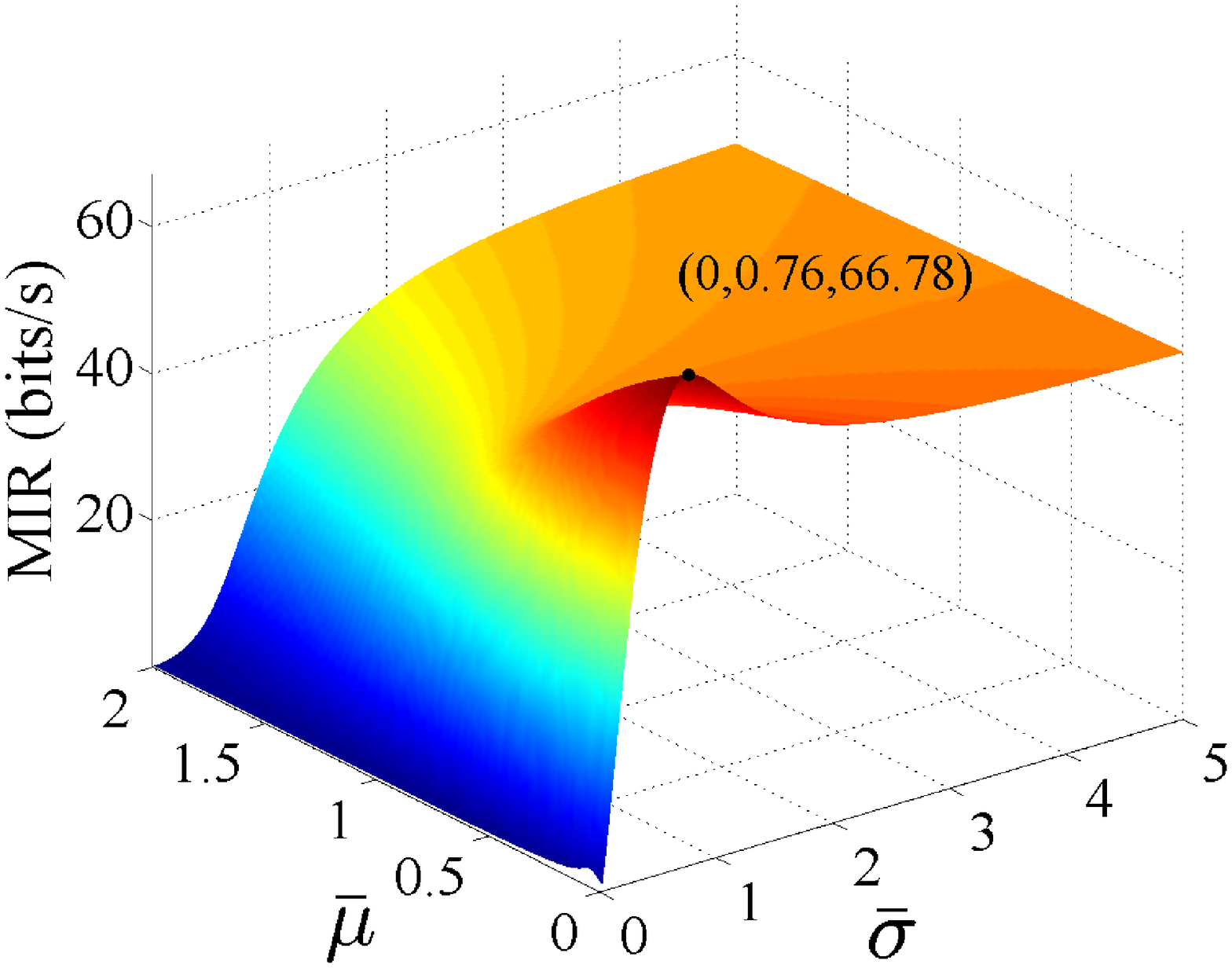}
    \label{ACh_accurate}
    }
        \subfigure[Upper bound]{
	\includegraphics[width=2.2in]{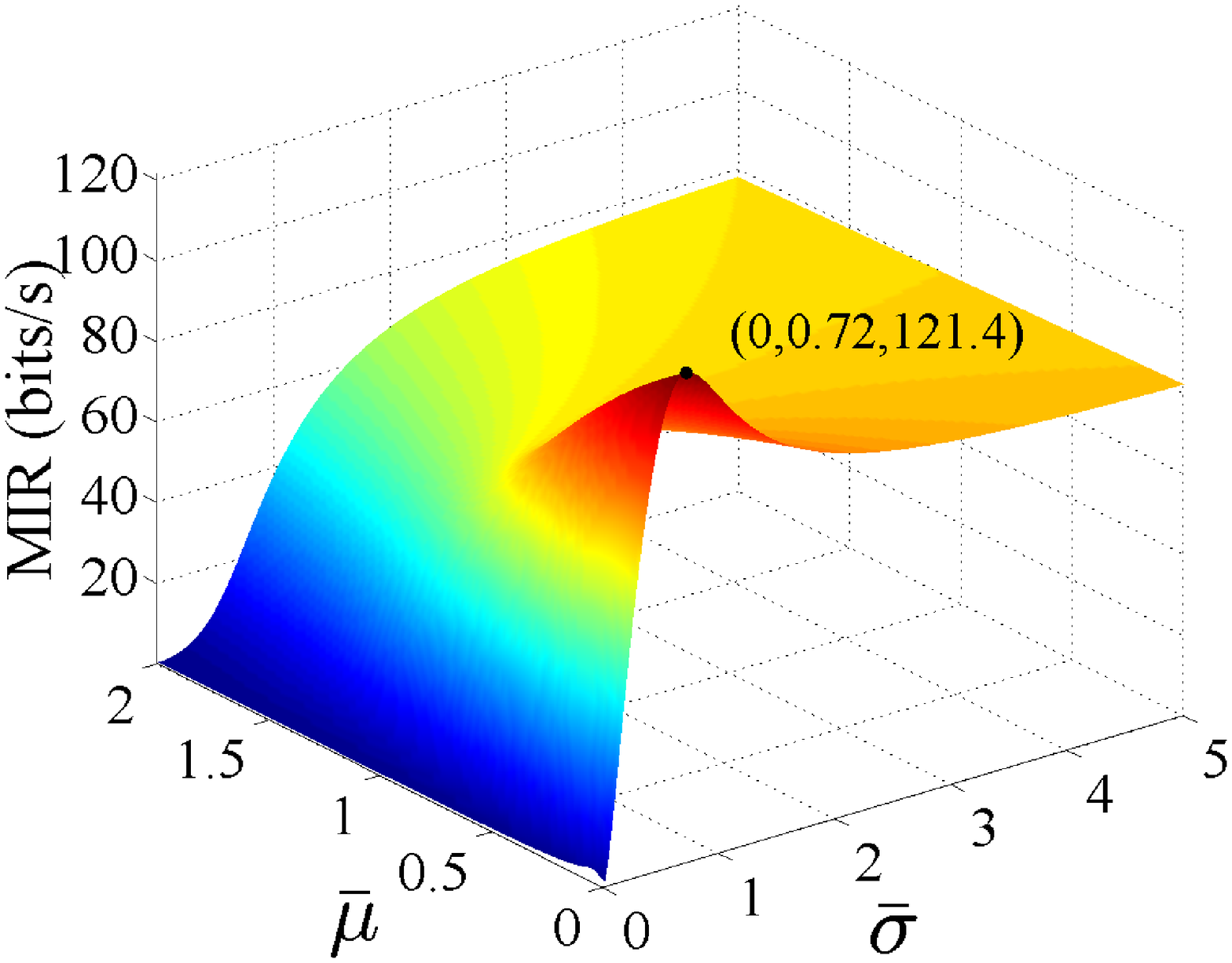}
	\label{ACh_upper_bound}
	}
	    \subfigure[Lower bound]{
	\includegraphics[width=2.2in]{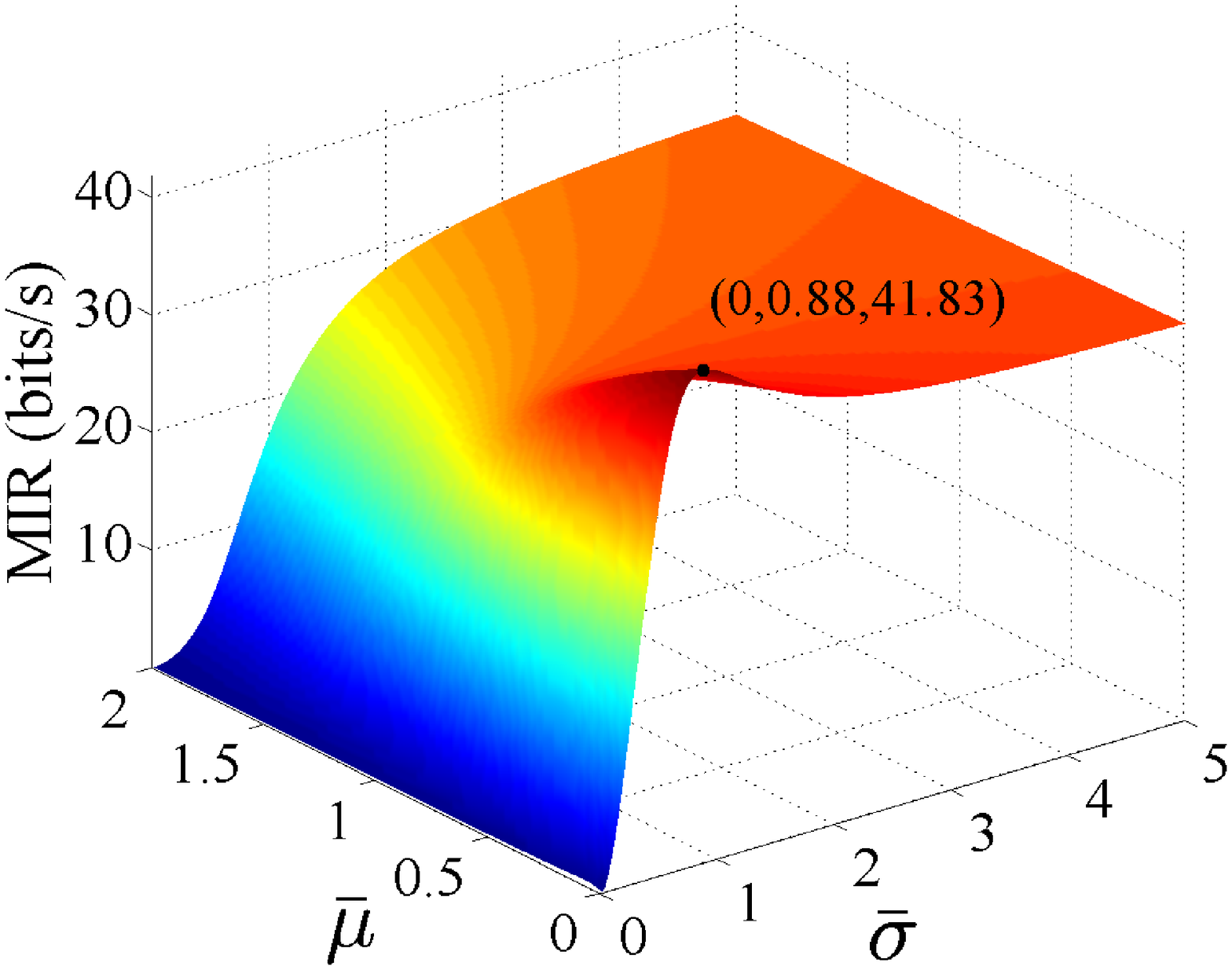}
	\label{ACh_lower_bound}
	}
 \vspace{-0.25cm}
    \caption{Numerical verification for the ACh receptor: the lower and upper bounds for $\mathop {\lim }\limits_{\Delta t \to 0} {\textstyle{{{\cal I}\left( {X;Y} \right)} \over {\Delta t}}}$, where it is assumed that the input $x$ follows the truncated Gaussian distribution with $x \in \left[ {2 \times 10^{-2},2} \right]$ and $s=2$.}
    \label{Numerical verification for the ACh receptor}
\end{figure*}

Fig.~\ref{Numerical verification for the ACh receptor} depicts the MIR and its bounds when simultaneously considering $\bar \mu$ and $\bar \sigma$ for the ACh receptor. Particularly, given the high calculation complexity from the derived bounds with $s=4$, its curves have been removed for the ACh receptor.
% Considering the significant performance of the approximately numerical solution in \eqref{Approximately Numerical Solution}, its curve has been removed for the ACh receptor.
By comparing Fig.~\ref{ACh_accurate} and Figs.~\ref{ACh_upper_bound}-\ref{ACh_lower_bound}, we can discover that the trends of the MIR and its bounds near-perfectly match, especially for the capacity-achieving values of $\bar \mu$ and $\bar \sigma$. Thus, the derived bounds can provide a possible range~for $\mathop {\lim }\limits_{\Delta t \to 0} {\textstyle{{{\cal I}\left( {X;Y} \right)} \over {\Delta t}}}$, while predicting its trend with varying parameters. For the receptor with a simple input distribution model, the input distribution reaching the capacity can be approximately calculated via \eqref{bound_July} to guide the system design.

\section{Conclusion}
\vspace{-0.1cm}
This paper brings new insights into the information theory analysis for the signal transduction channels. Specifically, we first derived an asymptotic continuous-time MIR for the signal transduction channel with the IID Gaussian input. To improve the practicality, the approximate numerical expression for the continuous-time MIR was given. Meanwhile, its lower and upper bounds for the considered MIR were deduced in closed-form. Finally, simulation results with the ChR2 and ACh receptors validated our analysis.
 
\bibliographystyle{IEEEtran} 
\bibliography{IEEEabrv, Capacity}

\end{document}